\documentclass[11pt,english]{article}
\usepackage{babel}
\usepackage{graphicx}
\usepackage{dcolumn}
\usepackage{bm}

\def\b{{\bf b}}

\def\F{{\bf F}}
\def\B{{\bf B}}
\def\n{{\bf n}}

\def\A{{\bf A}}
\def\r{{\bf r}}
\def\i{{\bf i}}
\def\j{{\bf j}}
\def\k{{\bf k}}
\def\p{{\bf p}}

\def\beq{\begin{equation}}
\def\eeq{\end{equation}}

\begin{document}

%\baselineskip=24truept

\title{Majorana spin-flip transitions in a magnetic trap}
\author{D.M.Brink and C.V. Sukumar\\
Department of Physics, University of Oxford\\
Theoretical Physics, 1 Keble Road, Oxford OX1 3NP}
\maketitle

%\noindent Date \today

\begin{abstract}
Atoms confined in a magnetic trap can escape by making spin-flip Majorana transitions due to a breakdown of the adiabatic approximation. Several papers have studied this process for atoms with spin $F = 1/2$ or $F= 1$. The present paper calculates the escape rate for atoms with spin $F > 1$.  This problem has new features because the perturbation $\Delta T$ which allows atoms to escape satisfies a selection rule $\Delta F_z = 0, \pm 1, \pm 2$ and multi-step processes contribute in leading order. When the adiabatic approximation is satisfied the leading order terms can be summed to yield a simple expression for the escape rate.

\end{abstract}

\section{Introduction}

This is a more complete version of a Brief Report published in Physical Review A {\bf 74} 035401 (2006). 
Magnetic traps for neutral atoms have many applications in atomic physics and quantum optics. For example they are used to confine and study Bose-Einstein condensates. Magnetic wave guides can transmit and manipulate atomic de Broglie waves. Atoms may be trapped in fine structure level or hyperfine levels depending on the nuclear spin. Alkali atoms which have a non-zero nuclear spin  are trapped in hyperfine states, while atoms of $^{52}$Cr (Schmidt et al \cite{Schmidt}), which have a zero nuclear spin but a large electronic spin are trapped in fine structure states. In this paper we focus on atoms trapped in hyperfine states, but the theory of fine structure trapping is essentially the same.

Atoms are confined only in certain Zeeman levels and spin-flip transitions cause them to be lost from a trap.
Thermal fluctuations in the surrounding apparatus can cause atoms to make spin-flip transitions to untrapped states. The theory of thermally induced losses  has been developed by Henkel et al \cite{Henkel} and has been studied  experimentally by Harber et al \cite{Harber}. Losses due to radio-frequency noise in the currents that form the microtrap have been studied by Leanhardt et al \cite{Leanhardt}. Other references can be found in these papers. Spin-flip transitions between Zeeman levels due to a breakdown of the adiabatic approximation can also cause atoms to escape from a trap.  These transitions were first studied in a one dimensional time-dependent model by Majorana \cite{Majorana} in (1932)  and are often called Majorana transitions. In many experiments Majorana losses  are much less important than those due to  environmental effects; but they do come into play in some experimental situations (Schmidt et al \cite{Schmidt}, Ott et al. \cite{Ott}).

The spin of an atom in a magnetic trap precesses about the direction of the local magnetic field with frequency $\omega_{prec}$. At the same time its center of mass oscillates in the field with frequency $\omega_{vib}$. In the adiabatic limit, when $\omega_{vib}<< \omega_{prec}$, the component $F_z$ of hyperfine spin $F$ along the direction of the local field is an approximate constant of the motion. Under these conditions the atom moves in an effective potential $V_{ad}(\r)= \mu_0 g F_z B(\r) $ where $B(\r)$ is the magnitude of the local magnetic field $\B (\r)$, $\mu_0 $ is the Bohr magneton and $g$ is the hyperfine g-factor. Spin states with $g F_z >0$ can be trapped near a minimum of $B(\r)$.

Corrections to the adiabatic approximation \cite{SB1} cause spin-flip a transition to a state with $g  F_z \leq 0$ and the atom escapes from the  trap.  The cases   $F=$ 1/2 or 1 were studied in  Sukumar and Brink \cite{SB1} and the transition rates for atoms to escape from a trap were calculated. Similar calculations have been made by other authors \cite{Gov}. In the present paper we extend the results of ref.\cite{SB1} to cases where the hyperfine spin $F > 1$ 

The perturbation terms $\Delta T_1$ and $\Delta T_2$ in eq.(\ref{DeltaT}), which cause a breakdown of the adiabatic approximation, are calculated in section 4 and the transition rate for spin flip transitions is calculated in section 5.
The dominant perturbations contribution of $\Delta T_1$ has a selection rule $\Delta F_z =0, \ \pm 1$ and, for $F>1$ while $\Delta T_2$ also allows transitions with $\Delta F_{z} = \pm 2$. When $F>1$ higher order perturbations and interference effects are important.

Explicit expressions for Majorana transition rates are given in section 5. They show how the rates depend on the vibrational frequency of an atom in a magnetic trap, the Zemann splitting and the spin of the initial state.

\section{The adiabatic Hamiltonian}

An atom with mass $m$, hyperfine spin $F$ and magnetic moment ${\bf \mu}$ moves in a magnetic field ${\bf B(r)}$. The interaction energy is 
\begin{equation}
V= - {\bf \mu}\cdot \B =  \mu_0 g{\bf F.B}
\end{equation}
where $\mu_0$ is the Bohr magneton and $g$ is a $g$-factor for the hyperfine state and the Hamiltonian for  the motion of the atom coupled with the dynamics of the spin is
\begin{equation}
H = \frac{\p^2}{2m}  + \mu_0 g{\bf F.B}= T + h(\r).
\end{equation}

The adiabatic method starts as in ref.\cite{SB1} by introducing  basis states $|F_z({\bf r})\rangle$ which are eigenstates of the component of $\F$ along the direction of the local magnetic field. They are obtained from the eigenstates   $|F_z\rangle$ of $F_z$ by an ${\bf r}$-dependent rotation
\begin{equation}
|F_z({\bf r})\rangle = U({\bf r})|F_z\rangle.
\end{equation}

The Hamiltonian in the adiabatic basis is
\begin{equation}
H^{\prime} = U^{-1}(\r)H U(\r)= \frac{\bf p^2}{2m} + V_{ad}(r) + \Delta T
\end{equation}
where $V_{ad}(r)$ is the adiabatic potential and $\Delta T $ is a correction coming from the  transformation to the adiabatic basis. The adiabatic approximation corresponds to neglecting  $\Delta T$ and corrections are obtained by treating it as a perturbation.
The analysis of ref.\cite{SB1} shows that $\Delta T = \Delta T_1 + \Delta T_2$ where
\begin{equation}
\Delta T_1 = \frac{1}{2m}[ {\bf A(r).p + p. A(r)}],\qquad 
 \Delta T_2 = \frac{1}{2m} A.A 
\label{DeltaT}
\end{equation}
where ${\bf A(r)} = -i\hbar U^{-1}\nabla U(\r)$.

Following Sukumar and Brink \cite{SB1} the unitary operator $U(\r)$ is chosen 
to be a rotation through 
 $\pi$ about an axis $\n(\r)$ where $\n$ is a unit vector which bisects the angle between the z-axis and the direction of ${\bf B(r) }$ of the field when the atom is at $\r$.
Hence $U(\r)=\exp(i\pi\n(\r)\cdot \F)$. If the atom moves then $n(\r)\rightarrow n(\r +\delta\r)$. 
To first order in  $\delta \n$  
\begin{equation}
U^{-1}\delta U= 2 i \n\times \delta \n\cdot \F.
\end{equation}
 The resulting expression for ${\bf A(r)}$ is
\begin{equation}
{\bf A(r)} = -i\hbar U^{\dagger}({\bf r}) [\nabla U({\bf r})]=
2 \hbar \n\times \nabla \n\cdot \F.
\label{A}
\end{equation}

In the adiabatic basis ${\bf F}$ has components  $F_z$ along the direction of the local field ${\bf B}$ and the adiabatic potential is 
\begin{equation}
V_{ad}(\r) =  \mu_0 g F_z  B(r) 
\label{Vad1}
\end{equation}
where $B(\r)$ is the magnitude of the magnetic field. 
 There is a set of adiabatic potential surfaces, each with its own value of $F_z$.
For simplicity we assume that $\mu_0 g >0$ so that states with $F_z >0$ can be trapped around the minimum of $B(\r ) $ at $\r = \r_m$. 
When $F_z \leq 0$ the adiabatic potential $V_{ad}$ has a maximum at $\r = \r_m$ and the states are not trapped. 

Trapped atoms in states with $F_z > 0$ can escape from the trap by making a transition to a state with $F_z \leq 0$ because of non-adiabatic effects.
The perturbation $\Delta T$ mixes states with different $F_z$ and can cause transitions out of the trap. The quantity ${\bf A(r)}$ is proportional  to the spin ${\bf F}$ so that the perturbation $\Delta T_1$ can mix states with $\Delta F_z =0, \ \pm 1$. The perturbation $\Delta T_2$ has a selection rule $\Delta F_z = 0, \ \pm 1, \ \pm 2$. 
Trapped atoms with $F_z = 0$ or $1$ can escape by making a single step transition. This was the case discussed in ref. \cite{SB1}. For atoms with $F_z > 1$  higher order effects must be included.

\section{ A simplified choice for the magnetic field}

The results in the last section hold for a general magnetic field. Many different devices for trapping neutral atoms have been developed. A typical configuration is the Ioffe-Pritchard trap \cite{Pritchard}. The fields are produced by current carrying coils or by permanent magnets. Near the center of the trap the magnetic field is approximately
\beq
\B \approx B_0 \k +B_{\rho}^{'}(x \i - y \j) + 
\frac{B_z^{''}}{2}\left(z^2 - \frac{x^2 + y^2}{2} \right) \k + ...
\label{field}
\eeq
and is characterized by an axial bias magnetic field $B_0$, an axial field curvature 
$B_z{''}$ and a radial quadrupole field gradient  $B_{\rho}^{'}$. 

Traps for ultracold atoms have been constructed on several different length scales and in many of these cases the field near the center of the trap can be approximated by eq.(\ref{field}).  The fields in conventional macro-traps, like the one used in the experiments of Harber et al \cite{Harber}, have a size on the scale of several centimeters. Typical values for the radial gradient and axial curvature are 50 G/cm and 100 G/cm$^2$ respectively. The fields in micro-traps are produced by current carrying wires on the surface of a microchip (Folman et al \cite{Folman}). Distances from the wire to the center of the trap are in the range 100-1000 $\mu$m with a radial field gradient in excess of  of 10,000 G/cm. 
Atom wave guides have a non-zero axial bias field and radial gradient but with an axial curvature $B_z^{''} =0$ so there is no confinement in the z-direction.

To simplify the calculations in this paper  the field ${\bf B(r)}$ is assumed to have a constant axial bias field $B_0$ and a radial field gradient $B_{\rho}^{'} = \lambda$. The  axial curvature is set equal to zero so that there is no confinement in the z-direction.  This can describe the field in an atomic wave guide with its axis parallel to the z-axis or  a simplified trap where the atom is restricted to move in the $(x,y)$ plane.
The spin has components $F_z$ along the direction of the local field ${\bf B}$ and the adiabatic potential is 
\begin{equation}
V_{ad} =  \mu_0 g F_z  B(r)=  
\mu_0 g F_z\sqrt{B_0^2+\lambda^2(x^2 +y^2)}.
\label{Vad2}
\end{equation}

When the uniform field $B_0$ is large enough the adiabatic potential can be approximated by a harmonic oscillator 
\begin{equation}
V_{ad}  \approx \mu_0 g F_z \left (B_0+\frac{\lambda^2}{2B_0}(x^2+y^2)\right).
\label{Vad3}
\end{equation}

The adiabatic potential depends on $F_z$ and both the oscillator frequency $\omega_{vib}$ and length parameter $b$ depend on $F_z$. They are given by
\[\omega = \omega_0\surd {F_z}, \qquad b^2 = b_0^2 /\surd {F_z}\]
where
\begin{equation}
\omega_0^2 = \frac{\mu_0 g  \lambda^2}{m B_0} \ , 
\qquad b_0^2 = \frac{\hbar}{m\omega_0}
= \sqrt{\frac{\hbar^2 B_0}{m\mu_0 g  \lambda^2}}.
\label{b0}
\end{equation}
We assume that $\mu_0 g >0$ so that states with $F_z >0$ are trapped in the adiabatic potential. When $F_z \leq 0$ then $V_{ad}$ is an inverted parabola and the states are not trapped.

The energy gap between the adiabatic surfaces at $r = 0$ is 
\beq
E_0 = \mu_0 g B_0 = \hbar \omega_{prec}
\eeq
where $\omega_{prec}$ is the precession frequency of the spin in the magnetic field at $r = 0$.
The adiabatic approximation is reliable when
\begin{equation}
\chi_0 = \frac{\omega_0}{\omega_{prec}} = \frac{\hbar \omega_0}{E_0} = 
\frac{\lambda^2 b_0^2}{B_0^2} = \sqrt{\frac{\hbar^2 \lambda^2}{m \mu_0 g B_0^3}}  << 1.
\label{chi0}
\end{equation}
This is equivalent to the adiabatic condition in \cite{SB1}. The parameters $\omega_0$, $B_0$, $E_0$ and $\chi_0$ depend only on the fields and not on the spin.

The initial state of the atom in the trap has spin component $F_z = F_{zi}$ oscillator length parameter $b_i$ and  oscillator frequency  $\omega_i$ . 
The energy of the atom in the initial state is 
\beq
E_i  = F_{zi}\mu_0 g B_0 + \hbar \omega_i \approx F_{zi} \omega_{prec}
\label{Ei}
\eeq
when $\chi_0 <<1$.

In the case of integer spin there the adiabatic potential is zero in the final state and the atom has a kinetic energy fixed by the energy conservation condition
\[
 E_i = F_{zi}\mu_0 g B_0 =  \hbar \omega_{prec} F_{zi}.
\] 

We see from the second last term in eq.(\ref{chi0}) that the condition that the frequency ratio $\chi_0$ is small ensures that the expansion (\ref{Vad3}) is a good approximation for the low states in the adiabatic potential. 

\bigskip

\section{Calculation of $\Delta T_1$ and $\Delta T_2$}

The spinor field $\A(\r)$ is given by eq.(\ref{A}) where the unit vector $\n$ is chosen as in  ref.(\cite{SB1}) 
\begin{equation}
\n=\beta x\i -\beta y\j + \alpha \k
\end{equation}
with
\begin{equation}
\alpha^2 = \frac{B_0 + B}{2 B}, \qquad \beta^2 = \frac{\lambda^2}{2B ( B_0 + B)}.
\label{alphabet}
\end{equation}
Substituting the expression for $\n\times \nabla \n\cdot \F$ into eq.(\ref{A}) gives
\begin{eqnarray}
\A(\r)	\ &=& \A_1(\r) + \A_2(\r)  + \A_3(\r)\qquad {\rm with} \\
\A_1(\r) &=& 2\hbar  \alpha \beta (F_y \nabla x + F_x \nabla y), \\
\A_2(\r) &=& 2\hbar \beta^2(y \nabla x - x \nabla y)F_z,\\
\A_3(\r) &=& 2\hbar (y F_x + x F_y) (\alpha\nabla \beta -\beta \nabla \alpha).
\end{eqnarray}
This equation for $\A(\r)$ corresponds to one given in ref.( \cite{SB1}). The explicit expression for 
$\beta^2$ is given in eq.(\ref{alphabet}) and related factors in the equations for $\A_1(\r)$ 
and $\A_3(\r)$ simplify to
\[\alpha \beta =\frac{\lambda}{2B},\qquad
(\alpha\nabla \beta -\beta \nabla \alpha)= \frac{\lambda^3 \r}{2 B^2(B + B_0)}.
\]

The eigenstates in the adiabatic potential are $|L,n,F_z \rangle$, where $L$ is the orbital angular momentum about the symmetry axis and $n$ is the number of nodes in the radial wavefunction.  The energy eigenvalues in the harmonic approximation are
\beq
E = \mu_0gF_z + \sqrt{F_z}\hbar \omega_0 (|L| + n +1).
\eeq
The matrix elements of $\Delta T$ conserve the quantity $L-F_z$. This can be seen explicitly by examining the contributions of the various terms in $\Delta T_1$ and $\Delta T_2$ and also, more generally, from the symmetry properties of  the magnetic interaction which is proportional to
\beq
\B \cdot \F = B_0 F_z + x F_x - y F_y = 
B_0 F_z + \frac{1}{2}(x_{+}F_{+} + x_{-}F_{-})
\eeq
where $x_{\pm} = x \pm i y$ and $F_{\pm} = F_x \pm iF_y$.

The terms $ \A_2\cdot \p, \ \A_1\cdot \A_1$ and $ \A_2\cdot \A_2$ in $\Delta T$ all commute with $F_z$ while $\A_2\cdot \A_3=0$. These terms do not contribute to transitions out of the trap. The leading contributions in the adiabaticity parameter $\chi_0$ come from the terms proportional to $\A_1\cdot \p$  and $\A_1\cdot \A_3$. They are 
\beq
V_1 =\frac{1}{2m}\A_1\cdot \p = \frac{\hbar \lambda}{2 m B}(p_x F_y  + p_y F_x )=
-i\frac{\hbar \lambda}{4 m B}(p_{+}F_{+} -p_{-}F_{-}),
\label{V1}
\eeq
and 
\beq
V_2 =\frac{2\hbar^2}{m}\frac{\lambda^4}{4B^3(B+B_0)}(xF_y +yF_x)^2) =
 \frac{\hbar^2}{m}\frac{\lambda^4}{8B^3(B+B_0)}(x_{+}F_{+} - x_{-}F_{-})^2
\label{V2}
\eeq
where $p_{\pm} = p_x \pm i p_y$.
The contribution of $V_1$ in second order and the contribution of 
$V_2$ in first order are both proportional to $\chi_0^{2}$. In the following $B$ is approximated by $B_0$.

\section{Transitions out of the trap}
Atoms with a g-factor $g>0$ and with $F_z >0$ are trapped in the adiabatic potential when the perturbation terms $\Delta T_1$ and $\Delta T_2$ are neglected. They can escape by making a transition to  states with $F_z \leq 0$ under the influence of the perturbations. The dominant terms in a perturbation series are $V_1 \propto\A_1\cdot \p /2m$ with selection rule 
$\Delta F_z =  \pm 1$ and $V_2 \propto \A_1\cdot \A_3/2m$ with selection rule $\Delta F_z = 0, \ \pm 2$. 
Trapped states with $F_z =  \ 1/2$ or $ \ 1$ can escape by a single step transition under the influence of $V_1$. This  case was discussed in ref. \cite{SB1}.

In the following section it is assumed that the initial spin state is the ground state in the adiabatic potential with $F_z = F_{zi}$, radial quantum number $n =0$ and angular momentum component $L_z = L =0$
\[
|L, n, F_z\rangle =  |0, 0, F_{zi}\rangle . 
\]
 The perturbation $\Delta T $ introduces admixtures of states with smaller values of $F_z$ and values of $L$ satisfying the selection rule $\Delta (F_z -L) =0$. The emission of the atom from the trap depends on the amplitudes of states with $F_z =2, \ 3/2, \ 1, \ 1/2 $ in the initial state. These components allow transitions to unbound states with $F_z\leq 0$ under the influence of the perturbation and the initial state  acquires a width $\Gamma$. When the adiabatic approximation is valid $\Gamma <\hbar \omega_i < \hbar \omega_{prec}$ where $\omega_i$ is the vibrational frequency in the initial adiabatic potential and $\omega_{prec}$ is the  precession frequency of the initial state. 

The orbital part of the initial state is an s-state harmonic oscillator wave function
\[
\phi_i(r) = \frac{1}{b_i\surd{\pi}}\exp\left(-\frac{r^2}{2b_i^2}\right)
\]
in the adiabatic limit.
The leading contributions to the transition rate out of the trap are due to the parts 
$V_1^{-}$ and $V_2^{-}$ of the operators $V_1$ and $V_2$ which decrease $F_z$. They are 
\[
V_1^{-} = i \frac{\hbar \lambda }{4mB_0} p_{-}F_{-}
= i \frac{\hbar \omega_0}{4}\surd{\chi_0}\frac{b_0}{\hbar }p_{-}F_{-} , 
\]
\[
V_2^{-} = \frac{\hbar^2}{16 m} \frac{\lambda^4}{B_0^4}x_{-}^2 F_{-}^2 =
\frac{\hbar \omega_0}{16}\chi_0^2 \frac{x_{-}^2}{b_0^2} F_{-}^2 \ ,
\]
where $b_0$ is defined in eq.(\ref{b0}) and $\chi_0$ in eq.(\ref{chi0}). 
The perturbation expression for the transition amplitude has contributions from $V_1^{-}$ and $V_2^{-}$ as well as interference terms. 

The initial state with $L=0, n=0 $ and $F_z=F_{zi}$ is denoted by $|\phi_i\rangle$, the final continuum state with $F_{zf} = 0$ or $-1/2$ by $| f \rangle$ and states with $L=-j$ and $F_z = F_{z i}-j$ by $| j,n \rangle$.
The number of positive values of $F_z$ is denoted by $p = F_{zi} - F_{zf} $.
  The amplitude $A$ for the decay out of the trap is given by the perturbation formula
(Messiah, Chapter XVI, section 6)
\[A = \sum_{j_1,n_1,j_2 n_2,..}\langle f | V^{-}| J_m,n_m\rangle 
\frac{\langle J_m,n_m | V^{-}| j_{m-1},n_{m-1}\rangle} {E_{j_m,n_m}-E_i} {\cdot \cdot \cdot} 
\]
\beq
\cdot \cdot \cdot \frac{\langle j_2,n_2 | V^{-}| j_1,n_1\rangle }{E_{2,n_2}-E_i}
\frac{\langle j_1,n_1 | V^{-}| 0\rangle }{E_{1,n_1}-E_i}
\label{perts}
\eeq
where $V^{-} = V_1^{-} + V_2^{-}$. Each matrix element in the series (\ref{perts}) contains either $ V_1^{-}$ or  $ V_2^{-}$, but not both, because they satisfy different selection rules for $\Delta L$. The general term in the series (\ref{perts}) contains $p_1$ matrix elements of $V_1^{-}$ and $p_2$ matrix elements of $V_2^{-}$ where
 $p_1 + 2p_2 = p$ is the 'number of perturbation steps;
\begin{eqnarray}
p & = & F_{zi} \qquad  \qquad \quad {\rm for \ integer \ spin} \nonumber \\
p & = & F_{zi} + 1/2 \qquad {\rm for \ half \ integer \ spin}.
\label{ps}
\end{eqnarray}

In equation (\ref{perts}) $E_i $ is the energy of the initial state and $E_{j_m,n_m}$ is the energy of the state $j_m,n_m$. They are given by
\[
E_i = E_0 F_{zi} +\hbar \omega_0 \sqrt{F_{zi}},\qquad 
E_{j_m,n_m} = E_0(F_{zi} - j) +(2n_m +1)\hbar \omega_0\sqrt{F_{zi}-j_m}
\]
and all depend on $F_z$ because the adiabatic potential depends on $F_z$. The expression for  $A$ is complicated because there is mixing of  many orbital states $n_i$ for each $j_i$.

There is a big simplification in the adiabatic limit. Then $\hbar \omega<< E_0$ and the energy denominators can be approximated by 
\[E_{j_m,n_m} -E_i \approx -j E_0
\]
and the sums over $n_1, n_2, ...$  in the perturbation formula for $A_1$ can be evaluated by closure. The structure of the final expression for the transition amplitude is
\beq
A \approx \hbar \omega_0 \left(\frac{\sqrt{\chi_0}}{4}\right)^p {\chi_0}^{p-1}
\langle F_{zf}|F_{-}^{p} | F_{zi}\rangle
\sum_{p_1 + 2p_2 =p}I_{p_1,p_2}(-1)^{p_2}N_{p,p_2}
\label{A1}
\eeq
where
\beq
I_{p_1,p_2} = \frac{b_0^{p_1 - 2p_2}}{(\hbar^{p_1})} 
\langle \phi_f|p_{-}^{p_1}x_{-}^{2p_2} | \phi_i \rangle
= \left(\frac{-1}{F_{zi}}\right)^{p_2}(-1)^{p_2}
\langle k_f, L_f|(b_0 p_{-}/\hbar)^p | \phi_i \rangle
\label{I12}
\eeq
and $k_f$ is the wave number of the final free article state.
The phase factor $(-1)^{p_2}$ in eq.(\ref{A1}) arises because all the energy denominators in (\ref{perts}) are negative, and the one in eq.(\ref{I12}) comes from the reduction of the  matrix element. They cancel and all the interference terms between $V_1^{-}$ and $V_2^{-}$ are constructive. The  $N_{p, p_2}$ are numerical combinatorial factors. For $p_2 =0$ and $p_2 =1$ they are given by
\[
N_{p,0} = \frac{1}{(p-1)!}, \qquad\qquad N_{p,1} = \frac{p}{2(p-2)!}.
\]
Some values for $p_2 = 2$ are given in the Appendix.

\subsection{Integer spin}
The expression for the transition rate out of the trap depends on whether the hyperfine spin of the atom is integral or half integral.
In the case of  integral spin the adiabatic potential is zero in the final state  and the orbital state of the atom is  a plane wave $\phi_f = (L)^{-1}\exp(i\k_f\cdot \r)$ normalised in a box with side $L$. The magnitude of $\k_f$ is fixed by energy conservation $k_f^2= 2m E_i/\hbar^2$. The orbital matrix elements can be evaluated and
\[I_{p_1,p_2} =  \left(\frac{1}{F_{z,i}}\right)^{p_2}b_0^p (k_x - ik_y)^p I_0(\k_f)
\]
where 
\beq
I_0(\k_f) = \frac{2\surd{\pi}b_i}{L} e^{-k_f^2 b_i^2/2}
\eeq
 is the overlap of the initial and final orbital states. 
Hence
\beq
A \approx \hbar \omega_0 \left(\frac{\sqrt{\chi_0}}{4}\right)^p \chi_0^{p-1}
\langle F_{zf}|F_{-}^{p} | F_{zi}\rangle b_0^p (k_x - ik_y)^pI_0(\k)C_p.
\eeq
 The factor 
\beq
C_p = \left( \sum_{p_2} \left(\frac{1}{F_{zi}}\right)^{p_2} N_{p,p_2}\right).
\label{Cp}
\eeq
Some values of $N_{p,p_2}/(F_{zi})^{p_2}$ and $C_p$ are given in Appendix A. The numbers show that the contribution of the perturbation $\Delta T_2 =\A.\A/2m $ to the transition rate is important. This is because the terms in the perturbation formuala add coherently.  As explained in section 5 $p$ is related to the spin component of the initial trapped state by $p = F_{zi}$ for integer spin and $p = F_{zi} + 1/2$ for half-integer spin.

The transition rate out of the trap is given by Fermi's Golden Rule
\beq
w_f = \frac{2\pi}{\hbar}|A|^2\rho_f\quad {\rm where }\quad \rho_f = 
\frac{m L^2}{2\pi \hbar^2 }
\label{wf}
\eeq
is the density of final states. The  square of the magnitude of the transition matrix element reduces to
\beq
|A|^2 \approx \frac{(\hbar \omega_i )^2}{8} \left(\frac{F_{zi}\chi_0^2}{8}\right)^{p-1}
|\langle F_{zf}|F_{-}^{p} | F_{zi}\rangle | ^2 
C_p^2 \frac{4\pi b_i^2}{L^2} e^{- k_f^2 b_i^2}
\eeq
where we have used
\beq
\omega_0^2 = \omega_i^2/F_{zi}\qquad {\rm and} \qquad \chi_0 b_0^2 k_f^2 = 2F_{zi}.
\label{omega}
\eeq
The first of the expressions in eq.(\ref{omega})  relates the frequency $\omega_i$ of the initial state to the standard frequency $\omega_0$ defined in eq.(\ref{b0}) and the second comes from the energy conservation condition
\beq
\frac{\hbar^2}{2m}k_f^2 = F_{zi} E_0 = F_{zi}\hbar \omega_0 \frac{1}{\chi_0}=
F_{zi} \frac {\hbar^2}{mb_0^2 \chi_0}.
\eeq
in combination with eqs.(\ref{b0}) and (\ref{chi0}). Substituting into eq.(\ref{wf}) gives an expression for the escape rate from the trap
\beq
w_f = \frac{\pi \omega_i}{2}\left( \frac{p \chi_0^2 }{8}\right)^{p-1}
|\langle F_{zf}|F_{-}^{p} | F_{zi}\rangle|^2 C_p^2 e^{-k_f^2 \b_i^2}.
\label{wf0}
\eeq
where $p = F_{zi}$ (eq. \ref{ps}).

In the case of integral spin the spin of the atom in the unbound final state is $F_{zf}=0$ and the angular momentum factor is
\beq
|\langle 0 | F_{-}^p | F_{zi}\rangle| ^2 = F(F +1)(F-1)F+2)..(F-p+1)(F+p)
= \frac{(F+p)!}{(F -p)!}
\eeq
where $F$ is the total spin of the atom.

The exponent in eq.(\ref{wf0}) can be written in a number of different ways by using the relations
\beq
k_f^2b_i^2 =  \frac{2 \surd{F_{zi}}}{\chi_0}= 2\frac{E_i}{\ \hbar \omega_i}. 
\label{kf}
\eeq
Equation (\ref{wf0}) shows that, for a given $k_f$, the transition rate is reduced when the 
width $b_i$ of the initial state decreases. The second term in eq.(\ref{kf}), $(k_f b_i)^2 = 2\surd{F_{zi}}\chi_0$ shows that
the escape rate is exponentially small when the adiabaticity parameter is small ($\chi_0 <<1$). For a given $\chi_0$ it is reduced by an increase in the spin component $F_{zi}$ of the initial state. The pre-exponential factor $\chi_0^p$ gives a further reduction in the escape  rate but the factors depending on  $F$ and $F_{zi}$  increase the escape rate.
 Using  $(k_f b_i)^2 = 2 E_i /\hbar \omega_i$   the escape rate is proportional to $\exp(-2 E_i/ \hbar \omega_i)$. It is small when the magnetic energy  is large compared with the vibrational energy in the initial state ($E_i >> \hbar \omega_i$).

Experimental papers normally give the axial bias field $B_0$ in Gauss (G) and the radial field gradient $B'_{\rho} = \lambda$ in G/cm. Expressions for $\omega_{prec}$ and $\omega_0$ in these units are
\beq
\omega_{prec} = 8.8 \times 10^6 g B_0 \ {\rm sec}^{-1}{\rm G}^{-1}, \qquad
\omega_0 = 74.6 \lambda \sqrt{\frac {g}{A B_0}} \ {\rm sec}^{-1}{\rm G}^{-1}. 
\eeq 
Values of the parameters vary a lot from one experiment to another. For example  a microtrap  described by Ott et al \cite{Ott} for $^{87}$Rb atoms operated in several different modes. In one mode $B_0 \approx 0.7$ G and $\lambda \approx 122$ G cm$^{-1}$ 
so that
$\omega_0 \approx 5 \times 10^3$ sec$^{-1}$and  $\omega_{prec} \approx 1.7 \times 10^6$. The ratio $\chi_0 \approx 2.9 \times 10^{-3} <<1$ and the  adiabatic approximation is expected to  be reliable. This situation is tyical of most Ioffe-Pritchard traps. 

There are cases where the  adiabatic parameter $\chi_0 \sim 1$. This happens if the radial field gradient is very large (cf \cite{Ott}) or if the axial bias field is very small (cf \cite{Schmidt}.
In the second mode of the trap described by Ott et al \cite{Ott}
$B_0 \approx 1$ G and $\lambda \approx 3 \times 10^5$ G cm$^{-1}$. In this case
$\omega_0 \approx 3.7 \times 10^6$ sec$^{-1}$, $\omega_{prec} \approx 2.4 \times 10^6$ and $\chi_0 \approx 1.5 $ and the adiabatic approximation is not valid.

The expression (\ref{wf0}) is interesting because the exponential factor  is proportional to probability density $P_0(k_f)$ for finding a final momentum  $k_f$ when the initial orbital state of the atom is the ground state in the adiabatic potential.  The expression can be generalized to the case of a general initial orbital state with momentum distribution $P(k)$;
\beq
w_f = \frac{\pi \omega_i}{2}\left( \frac{\chi_0^2 F_{zi}}{8}\right)^{p-1}
|\langle F_{zf}|F_{-}^{p} | F_{zi}\rangle|^2 C_p^2 \frac{\pi}{b_i^2}P(k_f).
\label{wf3}
\eeq
Here the ${\bar C}_p$ are different from the values in Table 1 but are still of the order of  unity. In this form the spin-flip transition rate is small when the probability for finding the momentum $\hbar k_f$ in the initial state is small.

\subsection{Half integer spin}

When the hyperfine spin is $F=1/2, \ 3/2 ...$ the final value of the spin projection is $F_{z,f}=-1/2$ and the final orbital wave function is a continuum state in an inverted parabolic adiabatic potential. The changes in the expression for the transition rate can be estimated by replacing the plane wave final state by a semiclassical approximation
\[
\phi_f (r) \approx \frac{1}{(1 + r^2/(k^2b_f^2)}\exp\left(\int_0^r\sqrt{k^2 + r^2/b_f^2}\right)
\]
which reduces to the plane wave near $r=0$. Here $k_f$ is the momentum of the atom in the final state at $r=0$ and $b_f$ is the oscillator length parameter of the inverted parabolic potential. If the integrals in the matrix elements are estimated by the method of stationary phase then the expression (\ref{wf0}) for the transition rate is replaced by

\beq
w_f = \frac{\pi \omega_i}{2}\left( \frac{p \chi_0^2 }{8}\right)^{p-1}
|\langle -1/2 |F_{-}^{p} | F_{zi}\rangle|^2 C_p^2 e^{- c k_f^2 \b_i^2}.
\label{wfhalf}
\eeq
where $p= F_{zi} + 1/2$ (eq.\ref{ps}) and 
\beq
c= \sqrt{2F_{zi}} \tan^{-1}(\sqrt{\frac{1}{2F_{zi}}}).
\label{c}
\eeq

The biggest effect of the factor $c$ is for a transition from an initial state with $F_{zi} =1/2$ to a final state with $F_{zf}= -1/2$ when $c = \pi/4$. For large $F_{zi}$ the factor $c \rightarrow 1$.
(There was a printing error in eq.(34) of ref.\cite{SB1}. A factor $\pi$ was omitted in the exponent).

\medskip
\noindent
The angular momentum factor is replaced by
\[
\langle -1/2| F_{-}^p | F_{zi}\rangle ^2 = 
(F+1/2)\frac{(F+p-1/2)!}{(F-p + 1/2)!} \ .
\]
With these substitutions eq.(\ref{wf0}) reduces to
\[
w_f = \frac{\pi \omega_i}{2} \exp\left( - \frac{\pi k_f^2 b_i^2}{4}\right)
\] 
when $F = F_{zi} =1/2$.

\bigskip
\section{Discussion and conclusions}

We have obtained approximate expressions for the decay rate of an atom from a magnetic trap due to breakdown of the adiabatic approximation. The results  hold for an arbitrary initial spin state of the atom. When the initial orbital state of the atom is the ground state in the  adiabatic potential the  decay rate in eq.(\ref{wf0}) is proportional to the vibration frequency of the initial state multiplied by an exponential factor 
(eqs. \ref{chi0}, \ref{wf0}, \ref{kf})
\beq
\exp(- c k^2_{f}b^2_i) =
\exp\left(- 2 c\frac{\sqrt{F_{zi}}}{\chi_0}\right)
=\exp\left(- 2c \frac{E_0 \sqrt{F_{zi}}}{\hbar \omega_0}\right).
\label{expf}
\eeq
The factor $c=1$ for integer spin and is given by eq.(\ref{c} for half-interger spin.
There is also a factor $(p\chi_0^2/8)^{p-1}$, where $p = F_{zi}$ for integer spin and $p = F_{zi} + 1/2$ for half-integer spin. This further reduces the transition rate in the adiabatic limit. These expressions generalize the results of \cite{SB1} where $p=1$.

Under normal experimental situations  Majorana transition rates are small  
because the adiabaticity parameter $\chi_0$ is small and the ratio of spin precession frequency is similar to the basic trap frequency $E_0/\hbar \sim \omega_0$ is large.
Under these conditions the exponential factor (\ref{expf}) is 
 very small. 
Eq.(\ref{chi0})shows that $\chi_0 \propto (B_0)^{-3/2}$.
Thus Majorana transitions can be seen when the bias field $B_0$ is reduced sufficiently to make 
$\chi_0 \sim 1$. The calculations in this paper use the harmonic approximation for the adiabatic potential. The arguments at the end of section 3 show that the harmonic approximation is valid whenever the adiabtic approximation is valid.

After the escape of the atom from the trap it has a kinetic energy equal to $E_i$. The transition rate is small when the final kinetic energy is much greater than the zero point center of mass energy in the initial state or when $k_f b_i>>1$. The argument leading to eq.(\ref{wf3}) shows that the transition rate is proportional to the probability density $P_0(k_f)$ for finding the final momentum $k_f$ in the  bound state in the initial adiabatic potential. It becomes small when $k_f$ is large. This form shows that decay rate out of the trap is largely determined by momentum matching between the initial and final state. At a high temperature  
$P_0(k_f)$ is replaced by a Boltzmann factor $\exp(- \hbar^2 k_f^2 /2m k_B T)$ as in ref. \cite{SB1}.

The analysis in section 4 shows that the perturbation terms proportional to $\A(\r).\p$ and $\A(\r). \A(\r)$ both contribute in leading order to decay out of the trap. Their contributions interfere constructively when the orbital state is the ground state in the initial adiabatic potential. The interference effects are contained in the factor $C_p^2$ (eq. \ref{Cp}).

The vibrational frequencies $\omega_{||}$ and $\omega_{\perp}$ of states in a 3-dimensional trap are determined by the axial field curvature $B_z''$ and the radial curvature $\lambda = B_{\rho}'$.  A detailed discussion of losses from a 3-dimensional trap due to a breakdown of the adiabatic approximation is more complicated than the case considered here; but  qualitative considerations based on eqs.(\ref{wf0}) and (\ref{wf3}) suggest that losses due to the parallel motion  will be negligable when $\omega_{||} <<\omega_{\perp}$.

\bigskip

\bigskip\bigskip

\section*{Appendix: The coefficients $N_{p,p_2}$}

This comes form the energy denominators in the perturbation formula. There are $p-1$ energy denominators and $N_{p,p_2} = 1/(p-1)!$ when $p_2 =0$. When $p_2=1$ one of the energy denominators is missing. This can happen in $p-1$ ways.

Simple counting gives
\[N_{p,0}= \frac{1}{(p-1)!}\quad {\rm when} \quad p_2=0\]
\[N_{p,1}= -\frac{p(p-1)}{2(p-1)!}\quad {\rm when} \quad p_2=1\]
The minus signs come because the energy denominators are all negative. In general it is 
$(-1)^{p_2}$. We could not find expressions for general $p$ and $p_2>1$. Numerical values for $p_2 =2$ and $p\leq 5$ are given in the following table.

\[ \left|
\begin{array}{cccccc}
N_{p,p_2} \ &p=1 \ & p=2 & p=3   & \ p=4 \ & p=5 \\
N_{p,0}   \ & 1 \ & 1 \ & 1/2 \ & 1/6 \   & 1/24 \\
N_{p,1}/p   \ & 0 \ & 1/2   & 1/2 \ & 1/4   \   & 1/12 \\
N_{p,2}/p^2   \ & 0 \ & 0   & 0 \   & 1/32 \   & 1/40 \\
C_p \      & 1 \ & 3/2 & 1 \   & 43/96   & 3/20 
\label{Table1}
\end{array}
\right| \]
\begin{center}
Table 1, Values of $N_{p,0}$, $N_{p,1}$, $N_{p,2}$ and $C_p$.
\end{center}

\end{document}